\pdfoutput=1

\documentclass[11pt,twoside,a4paper,cmspaper,final,collab]{cms-tdr}
\def\svnVersion{344468}\def\cmsCernNoTag{CERN-PH-EP/2015-307}\def\cmsCernDate{\today}\def\cmsMessage{Published in Physics Letters B as \href{http://dx.doi.org/10.1016/j.physletb.2016.05.033}{\doi{10.1016/j.physletb.2016.05.033.}}}

\begin{document}\cmsNoteHeader{SUS-14-021}

\hyphenation{had-ron-i-za-tion}
\hyphenation{cal-or-i-me-ter}
\hyphenation{de-vices}
\RCS$Revision: 337344 $
\RCS$HeadURL: svn+ssh://svn.cern.ch/reps/tdr2/papers/SUS-14-021/trunk/SUS-14-021.tex $
\RCS$Id: SUS-14-021.tex 337344 2016-04-06 21:17:44Z alverson $
\newlength\cmsFigWidth
\ifthenelse{\boolean{cms@external}}{\setlength\cmsFigWidth{0.49\textwidth}}{\setlength\cmsFigWidth{0.6\textwidth}}
\ifthenelse{\boolean{cms@external}}{\providecommand{\cmsLeft}{top\xspace}}{\providecommand{\cmsLeft}{left\xspace}}
\ifthenelse{\boolean{cms@external}}{\providecommand{\cmsRight}{bottom\xspace}}{\providecommand{\cmsRight}{right\xspace}}
\providecommand{\NA}{\text{---}\xspace}
\newcommand{\mass}[1]{\ensuremath{m(#1)}\xspace}
\newcommand{\Iabs}{\ensuremath{I_\text{abs}}}
\newcommand{\Irel}{\ensuremath{I_\text{rel}}}
\newcommand\fermion{\ensuremath{\cmsSymbolFace{f}}\xspace}
\newcommand{\mt}{\ensuremath{m_\mathrm{T}}\xspace}
\newcommand{\mtautau}{\ensuremath{m_{\tau\tau}}\xspace}
\newcommand{\CT}{\ensuremath{C_\mathrm{T}}\xspace}
\newcommand{\LT}{\ensuremath{L_\mathrm{T}}\xspace}
\newcommand{\Wjets}{{\ensuremath{\PW\text{+jets}}}\xspace}
\newcommand{\Wpjets}{{\ensuremath{\PWp\text{+jets}}}\xspace}
\newcommand{\Wmjets}{{\ensuremath{\PWm\text{+jets}}}\xspace}
\newcommand{\Zjets}{{\ensuremath{\cPZ\text{+jets}}}\xspace}
\newcommand{\Zg}{\ensuremath{\cPZ/\gamma^*}\xspace}
\newcommand{\Zgjets}{\ensuremath{\cPZ/\gamma^*{+\text{jets}}}\xspace}
\newcommand{\Nb}{\ensuremath{N_\cPqb}\xspace}
\newcommand{\Nbsoft}{\ensuremath{N_\cPqb^\text{soft}}\xspace}
\newcommand{\Nbhard}{\ensuremath{N_\cPqb^\text{hard}}\xspace}
\newcommand{\DM}{\ensuremath{\Delta m}\xspace}
\newcommand{\ttl}{\ensuremath{\ttbar(1\ell)}\xspace}
\newcommand{\ttll}{\ensuremath{\ttbar(2\ell)}\xspace}
\newcommand{\VV}{\ensuremath{\cmsSymbolFace{VV}}\xspace}
\newcommand{\sTopsTop}{\ensuremath{\PSQt\,\PSQt}\xspace}
\newcommand{\dxy}{\ensuremath{d_\mathrm{xy}}\xspace}
\newcommand{\dz}{\ensuremath{d_\mathrm{z}}\xspace}

\providecommand{\sLepton}{\ensuremath{\widetilde{\ell}}\xspace}
\providecommand{\sLr}{\ensuremath{\widetilde{\ell}_\cmsSymbolFace{R}}\xspace}
\providecommand{\sLl}{\ensuremath{\widetilde{\ell}_\cmsSymbolFace{L}}\xspace}

\newlength{\vstop}
\settoheight{\vstop}{\sTop}
\newcolumntype{.}{D{.}{.}{-1}}
\newcolumntype{x}{D{,}{\,\pm\,}{-1}}
\newcolumntype{y}{D{,}{.}{3.7}}
\newcolumntype{z}{D{,}{.}{2.7}}
\newcolumntype{Z}{D{,}{.}{1.7}}
\cmsNoteHeader{SUS-14-021}
\title{Search for supersymmetry in events with soft leptons, low jet multiplicity, and missing transverse energy in proton-proton collisions at $\sqrt{s}=8$\TeV}

\date{\today}

\abstract{
Results are presented from a search for supersymmetric particles in scenarios with small mass splittings.
The data sample corresponds to 19.7\fbinv of proton-proton collisions recorded by the CMS experiment at $\sqrt{s}=8$\TeV.
The search targets top squark (\PSQt) pair production in scenarios with mass differences $\Delta m = m(\PSQt) - m(\PSGczDo)$ below the \PW-boson mass and with top-squark decays in the four-body mode ($\PSQt \to \PQb\ell\nu\PSGczDo$), where the neutralino (\PSGczDo) is assumed to be the lightest supersymmetric particle (LSP).
The signature includes a high transverse momentum (\pt) jet associated with initial-state radiation, one or two low-\pt leptons, and significant missing transverse energy.
The event yields observed in data are consistent with the expected background contributions from standard model processes.
Limits are set on the cross section for top squark pair production as a function of the \PSQt and LSP masses.
Assuming a 100\% branching fraction for the four-body decay mode, top-squark masses below 316\GeV are excluded for $\Delta m=25$\GeV at 95\% CL.
The dilepton data are also interpreted under the assumption of chargino-neutralino production, with subsequent decays to sleptons or sneutrinos.
Assuming a difference between the common \PSGcpDo{}/\PSGczDt mass and the LSP mass of 20\GeV and a $\tau$-enriched decay scenario, masses in the range $m(\PSGcpDo)<307$\GeV are excluded at 95\% CL.
}

\hypersetup{%
pdfauthor={CMS Collaboration},%
pdftitle={Search for supersymmetry in events with soft leptons, low jet multiplicity, and missing transverse momentum in proton-proton collisions at sqrt(s) = 8 TeV},%
pdfsubject={CMS},%
pdfkeywords={CMS, physics, supersymmetry}}

\maketitle
\section{Introduction}\label{sec:Introduction}

The main objectives of the CERN LHC programme include searches for new physics, in particular supersymmetry (SUSY)~\cite{SUSY0,SUSY1,SUSY2,SUSY3,SUSY4}, one of the most promising extensions of the standard model (SM) of particle physics.
Supersymmetric models can offer solutions to several shortcomings of the SM, in particular those related to the mass hierarchy of elementary particles~\cite{Witten:1981nf,Dimopoulos:1981zb} and to the presence of dark matter in the universe.

Supersymmetry predicts superpartners of SM particles (sparticles) whose spins differ by one-half unit with respect to their SM partners.
In SUSY models with $R$-parity~\cite{Farrar:1978xj} conservation, sparticles are pair-produced and their decay chains end with the lightest supersymmetric particle (LSP).
In many of these models the lightest neutralino (\PSGczDo) takes the role of the LSP and, being neutral and weakly interacting, would match the characteristics required of a dark matter candidate.
The LSPs would remain undetected and yield a characteristic signature of high missing transverse momentum, the magnitude of which is referred to as \ETmiss.

In this paper we investigate the production of supersymmetric particles in a scenario in which the mass splitting between the next-to-lightest SUSY particle (NLSP) and the LSP is small, which is referred to as compressed SUSY.
In this case, the events would escape classical search strategies because of the low transverse momenta (\pt) of the decay products of the NLSP.
Signal events can still be distinguished from SM processes if a high-\pt jet from initial-state radiation (ISR) leads to a boost of the sparticle pair system and enhances the amount of \ETmiss, while the other decay products typically remain soft.
In the signal scenarios studied in this paper, SUSY particles can decay leptonically, and the presence of low-\pt leptons can be used to discriminate further against otherwise dominant SM backgrounds, such as multijet production and \Zjets events with invisible \cPZ\ boson decays.

\begin{figure}
  \centering
\includegraphics[width=0.23\textwidth]{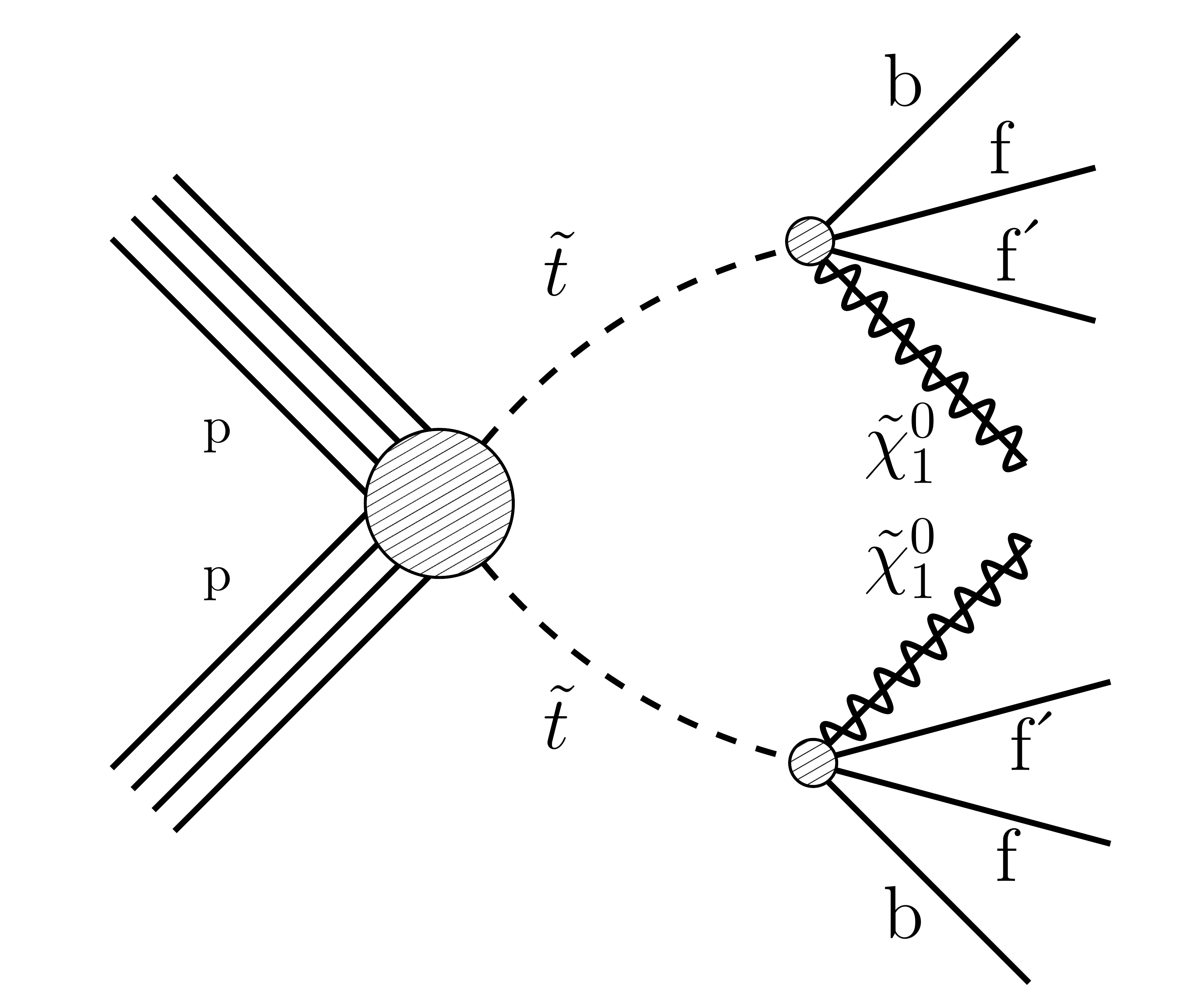} \hfil
\includegraphics[width=0.23\textwidth]{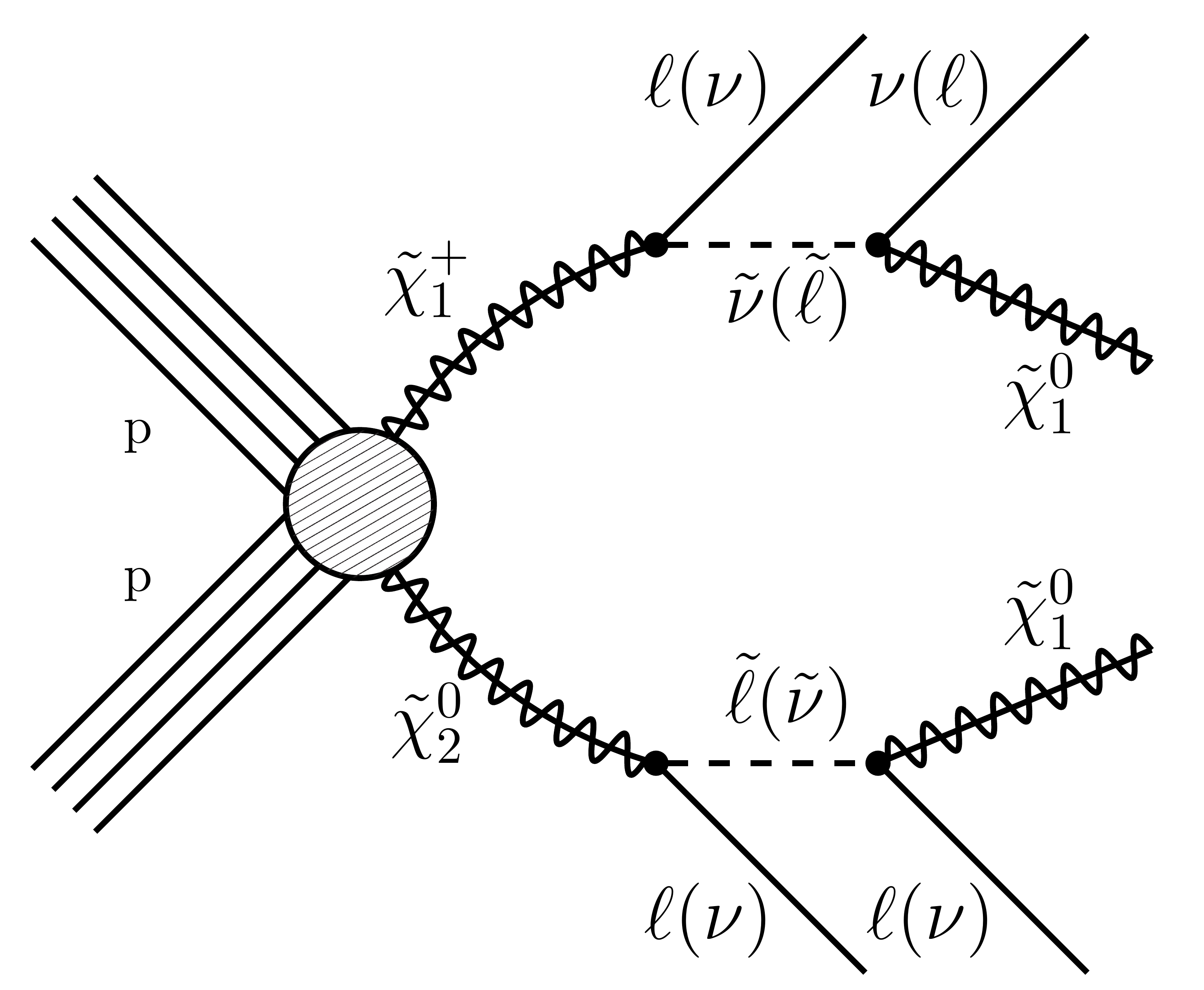}
\caption{Signal models for top squark pair production with subsequent four-body decays (left), and chargino-neutralino pair production with decays via sleptons and sneutrinos (right).
Antiparticle labels are suppressed.
The ISR jet used in the analysis is not shown in these diagrams.
}\label{fig:models}
\end{figure}

SUSY models with light top squarks (\sTop) are well motivated as they control the dominant correction to the Higgs boson mass and thereby preserve ``naturalness''~\cite{LightStop1,LightStop2,Dimopoulos:1981zb,Witten:1981nf,Dine:1981za,Dimopoulos:1981au,Sakai:1981gr,Kaul:1981hi}.
SUSY scenarios with mass splittings of 15--30\GeV between the top squark and the LSP are especially interesting because they would lead, through \sTop-\PSGczDo co-annihilation, to the observed cosmological abundance of dark matter~\cite{Coannihilation}.
For mass differences below the \PW-boson mass, top squarks could undergo either a two-body decay (such as $\sTop \to \cPqc \PSGczDo$) or a four-body decay ($\sTop \to \cPqb \fermion \fermion' \PSGczDo$, where $\fermion \fermion'$ represents a pair of quarks or leptons), as shown in the left panel of Fig.~\ref{fig:models}, with branching fractions and kinematic properties that depend on details of the model~\cite{Delgado:2012eu,Grober:2014aha}.
The search strategy based on the presence of an ISR jet has been used to search for the two-body decay in a monojet topology by the CMS Collaboration~\cite{Khachatryan:2015wza}, and for both decay modes by the ATLAS Collaboration~\cite{Aad:2014kra,Aad:2014nra,Aad:2015pfx}.
In this paper we assume that other SUSY particles are decoupled and that the four-body decay proceeds exclusively via virtual SM particles.

Final states with a hard ISR jet, high \ETmiss, and one or more charged leptons can also occur in the production of chargino-neutralino pairs in compressed SUSY models~\cite{Giudice:2010wb,Schwaller:2013baa,Han:2014kaa}.
A model of pair-production of the lightest chargino (\PSGcpDo) with the second-lightest neutralino (\PSGczDt) is shown in Fig.~\ref{fig:models}~(right).
Decay chains could proceed via intermediate sleptons or sneutrinos and give rise to final states with one or three charged leptons.
In this model, \PSGcpDo and \PSGczDt are assumed to be almost degenerate and are assigned a common mass \mass{\PSGc}.
In general, the same signature can arise from the production of heavy particles whose decay chains contain undetected, slightly lighter particles plus leptons.
Previous LHC results for the model of electroweak production described above and mass splittings below \mass{\cPZ} can be found in Refs.~\cite{Khachatryan:2014qwa,Aad:2014nua,Aad:2014yka,Khachatryan:2015kxa,Aad:2015eda}, where the last two references also report an alternative approach based on the vector-boson fusion topology.

In this paper we describe a search for pair production of top squarks with subsequent four-body decays via virtual top quarks and \PW\ bosons in events with a high-\pt jet, \ETmiss, and one or two soft leptons, corresponding to signal events with a leptonic decay of at least one of the virtual \PW\ bosons.
The single-lepton topology offers the second-highest branching fraction after the purely hadronic mode.
In this channel we consider only muons, which can be efficiently reconstructed and identified with transverse momenta as low as 5\GeV.
For the dilepton topology we require a second lepton (electron or muon) of opposite charge.
The single and double electron final states are not used because they have reduced sensitivity compared to the muon channels due to the higher \pt thresholds required for electrons.
In addition, selected events are required to have an energetic jet compatible with the ISR signature, at most one additional jet of moderate to high \pt, no hard leptons, and a significant amount of \ETmiss.
The dominant SM backgrounds to this search are pair production of top quarks, \PW\ boson or \Zg production in association with jets, and diboson (\VV) production.
Their contributions to the signal region (SR) are estimated by correcting the predictions from simulation using the event yields observed in several control regions (CRs) in data.
Data are also used to validate this procedure and to derive systematic uncertainties.

The results of the dilepton search are also interpreted in terms of the model of \PSGcpDo-\PSGczDt\ pair production discussed above.
For small  $\PSGcpDo-\PSGczDo$ mass splittings, the leptons in the final state would be soft and therefore  within the signal region of the dilepton search.

\section{Detector description and event reconstruction}\label{sec:objects}

The CMS detector has been described in detail in Ref.~\cite{ref:CMS}.
Its central feature is a superconducting solenoid that provides a homogeneous field of 3.8\unit{T} in a volume containing a silicon pixel and strip tracker, a lead tungstate crystal electromagnetic calorimeter, and a brass and scintillator hadron calorimeter.
Muons are measured in gas-ionization chambers embedded in the steel flux-return yoke surrounding the solenoid.
The acceptance of the silicon tracker and the muon systems extends to pseudorapidities of $\abs{\eta} < 2.5$ and $<$2.4, respectively.
The barrel and endcap calorimeters cover the range $\abs{\eta}<3.0$ and are complemented by extensive forward calorimetry.
Events are selected for further analysis by a two-tier trigger system that uses custom hardware processors to make a fast initial selection, followed by a more detailed selection executed on a dedicated processor farm.

The measurement of jets and \ETmiss is based on candidates reconstructed by the particle-flow (PF) algorithm~\cite{CMS-PAS-PFT-09-001,CMS-PAS-PFT-10-001}, which identifies leptons, photons, and charged and neutral hadrons by combining information from all subdetectors.
The PF candidates are clustered into jets by using the anti-\kt algorithm~\cite{Cacciari:2008gp} with a distance parameter of 0.5.
Jets are required to have $\pt > 30\GeV$ and $\abs{\eta} < 4.5$, and to pass loose quality criteria~\cite{CMS-PAS-JME-10-003} based on the energy fractions associated with electromagnetically or hadronically interacting charged or neutral particles.
The negative vector sum of the transverse momenta of the PF candidates defines the value of \ETmiss and the corresponding direction.
Jet energies and \ETmiss are corrected for shifts in the energy scale, contributions from additional, simultaneous proton-proton collisions (pileup), and residual differences between data and simulation~\cite{CMS-PAS-JME-13-004,Khachatryan:2014gga}.
Jets originating from \cPqb\ quarks are identified (``tagged'') using the combined secondary vertex algorithm~\cite{Chatrchyan:2012jua,CMS-PAS-BTV-13-001} at a working point corresponding to an efficiency of about 70\% and a misidentification probability for light-quark jets of about 1\%.
Hadronic decays of $\tau$ leptons are identified using the hadrons-plus-strips algorithm~\cite{Chatrchyan:2012zz}.

Muons and electrons are required to have \pt above 5 and 7\GeV, respectively.
In the single-muon search, the lepton acceptance is restricted to $\abs{\eta}< 2.1$, while in the dilepton search, this limit is tightened to 1.5 for both electrons and muons.
Standard loose identification requirements~\cite{Khachatryan:2015hwa,Chatrchyan:2012xi} are applied to reduce the background from nonprompt (NPR) leptons produced in semileptonic hadron decays and from jets showing a lepton signature.
Further background reduction is achieved by requiring the leptons to be isolated.
The absolute isolation \Iabs\ is computed by summing the transverse momenta of PF candidates, except that of the lepton, in a cone of size $\DR<0.3$ around the lepton direction, where $\DR \equiv \sqrt{\smash[b]{(\Delta\phi)^2+(\Delta\eta)^2}}$ and $\phi$ is the azimuthal angle measured in radians.
The energy in the isolation cone is corrected for the effects of pileup.
The relative isolation \Irel\ is obtained by dividing \Iabs\ by the \pt of the lepton.
The details of the isolation requirements differ between the single-lepton and dilepton topologies due to differences in the dominant backgrounds and the purities.
They are described in Sections~\ref{sec:singlelepton} and \ref{sec:dilepton}.
\section{Samples and event preselection}\label{sec:samples}

The data sample comprises proton-proton collisions recorded in 2012 at a centre-of-mass energy of 8\TeV and corresponds to an integrated luminosity of 19.7\fbinv.
The search uses events passing one of several online \ETmiss selections.
These triggers evolved over the data-taking period and required either $\ETmiss > 120\GeV$, where \ETmiss is reconstructed from the energy deposited in the calorimeters, or $\ETmiss > 95\GeV$ and a jet with $\pt > 80\GeV$ and $\abs{\eta}<2.6$, where both objects are reconstructed using the PF algorithm.
In the second part of the data-taking period, the threshold on \ETmiss was raised from 95 to 105\GeV.
Control samples were collected based on a single-muon trigger with a \pt threshold of 24\GeV.

Simulated Monte Carlo (MC) samples of SM background events are produced by using several generators.
Single and pair production of top quarks are simulated by using the \POWHEG~1.0~\cite{powheg} program.
Simulations of multijet and diboson events are done with \PYTHIA~6.4~\cite{Sjostrand:2006za}.
The generation of all other relevant samples, in particular \Zg processes, \Wjets events, and \ttbar production in association with a \PW, \Z, or Higgs boson, is performed with the \MADGRAPH~5.1~\cite{madgraph} generator.
Alternative samples of \ttbar and diboson events are also produced using \MADGRAPH to investigate possible systematic differences, which are found to be insignificant in the context of the analyses described in this paper.
All samples generated with \MADGRAPH or \POWHEG are passed to \PYTHIA~6.4 with the Z2* tune~\cite{Chatrchyan:2011id} for hadronization and showering.
The detector response is simulated with the \GEANTfour~\cite{GEANT4} program.
Finally, all events are reconstructed with the same algorithms as the ones used for data.
Pileup events are included in the simulation and all samples are reweighted to match the distribution of the average number of these events in data.

The signal simulation for \sTop\ pair production is done on a grid in the \sTop--\PSGczDo mass plane with \mass{\sTop} ranging from 100--400\GeV in steps of 25\GeV,  and $\DM \equiv \mass{\sTop} - \mass{\PSGczDo}$ ranging from 10--80\GeV in steps of 10\GeV.
The production of top-squark pairs with up to two additional jets and the four-body decays of the top squarks are generated with \MADGRAPH.
The decays are forced to proceed only through virtual SM particles.
Chargino-neutralino pair production is also modelled with \MADGRAPH, while their decays are generated with \PYTHIA.
We assume a bino-like LSP and wino-like \PSGczDt\ and \PSGcpDo\ in order to allow a direct comparison with Ref.~\cite{Khachatryan:2014qwa}.
A range in the common gaugino mass of 100--400\GeV is covered with steps of 20\GeV, maintaining a fixed mass difference of 20\GeV above the \PSGczDo.
As for the background samples, the generation steps for both signal models are followed by hadronization and showering in \PYTHIA.
For the signal samples, the modelling of the detector response is performed with the CMS fast simulation program \cite{Abdullin:2011zz}.
Differences in the efficiencies of the lepton selection and the \cPqb-jet identification between the fast and the detailed \GEANTfour simulation are corrected by using scale factors.
Deficiencies in the modelling of ISR in the simulation \cite{Chatrchyan:2013xna} are corrected by applying a weight as a function of the \pt of the recoiling system.

The effects of residual differences between data and simulation are taken into account in the analysis.
The systematic uncertainty related to possible variations in the jet energy scale~\cite{CMS-PAS-JME-13-004} is evaluated by a coherent change of all jet energies, which is also propagated to \ETmiss.
The jet energy resolution in simulation is found to be slightly better than in data~\cite{CMS-PAS-JME-13-004}.
To compensate for this effect, the energies of simulated jets are smeared and a corresponding systematic uncertainty is assigned.
Simulation is corrected for differences in the efficiencies of the reconstruction of leptons, and of the identification of leptons~\cite{Khachatryan:2015hwa,Chatrchyan:2012xi} and \cPqb\ jets~\cite{Chatrchyan:2012jua,CMS-PAS-BTV-13-001} with respect to the values measured in data.
The corresponding uncertainties are propagated to the final results.

The first step in the event selection is designed to match the online requirements and to serve as a common basis for the analysis in both channels.
It is guided by the general characteristics of signal events.
The leading jet of each event is considered as an ISR jet candidate.
It is required to pass tighter jet identification criteria and to fulfil $\pt>110\GeV$ and $\abs{\eta}<2.4$.
Since jets resulting from \sTop decays are soft, and no jets are expected from \PSGczDt or \PSGcpDo decays, at most one additional jet with $\pt>60\GeV$ is accepted.
At least one identified muon with $\pt>5\GeV$ and $\abs{\eta}<2.1$ must be present.
Finally, a requirement of $\ETmiss>200\GeV$ is imposed.
By using a control sample collected with the single-muon trigger, the signal triggers are found to be fully efficient after these preselection criteria are applied.

\section{Search in the single-lepton channel}\label{sec:singlelepton}

The single-lepton topology is selected by requiring at least one muon within the acceptance described in the previous sections.
Events are rejected if an electron, a $\tau$ lepton, or an additional muon with $\pt > 20\GeV$ is present.
To avoid strong variations of the muon selection efficiency with \pt, a combined isolation criterion, $\Iabs < 5\GeV$ or $\Irel < 0.2$, is used, equivalent to a transition from an absolute to a relative isolation requirement at $\pt = 25\GeV$.
The impact parameters of the muon with respect to the primary collision vertex in the transverse plane, \dxy, and longitudinal direction, \dz, are required to be smaller than 0.02 and 0.5\unit{cm}, respectively.
The primary vertex is chosen as the one with the highest sum of $\pt^2$ of its associated tracks.
Furthermore, requirements are imposed on \ETmiss and on the scalar sum of the transverse momenta of all jets, \HT.
Since these two observables are correlated, a simultaneous selection is applied by using the combined variable $\CT \equiv \min(\ETmiss,\HT-100\GeV)$.
To match the preselection, $\CT > 200\GeV$ is required.
Background from SM dijet and multijet production is suppressed by requiring the azimuthal angle between the momentum vectors of the two leading jets to be smaller than 2.5\unit{rad} for all events with a second hard jet of $\pt>60\GeV$.
According to simulation, the remaining sample is dominated by \Wjets and, to a lesser extent, by \ttbar production with a single prompt lepton in the final state.
Therefore, we use the transverse mass \mt~\cite{Arnison:1983rp} computed from the transverse components of the muon momentum and the \ETmiss vector as a discriminant.

\begin{figure}
  \centering
  \includegraphics[width=.49\textwidth]{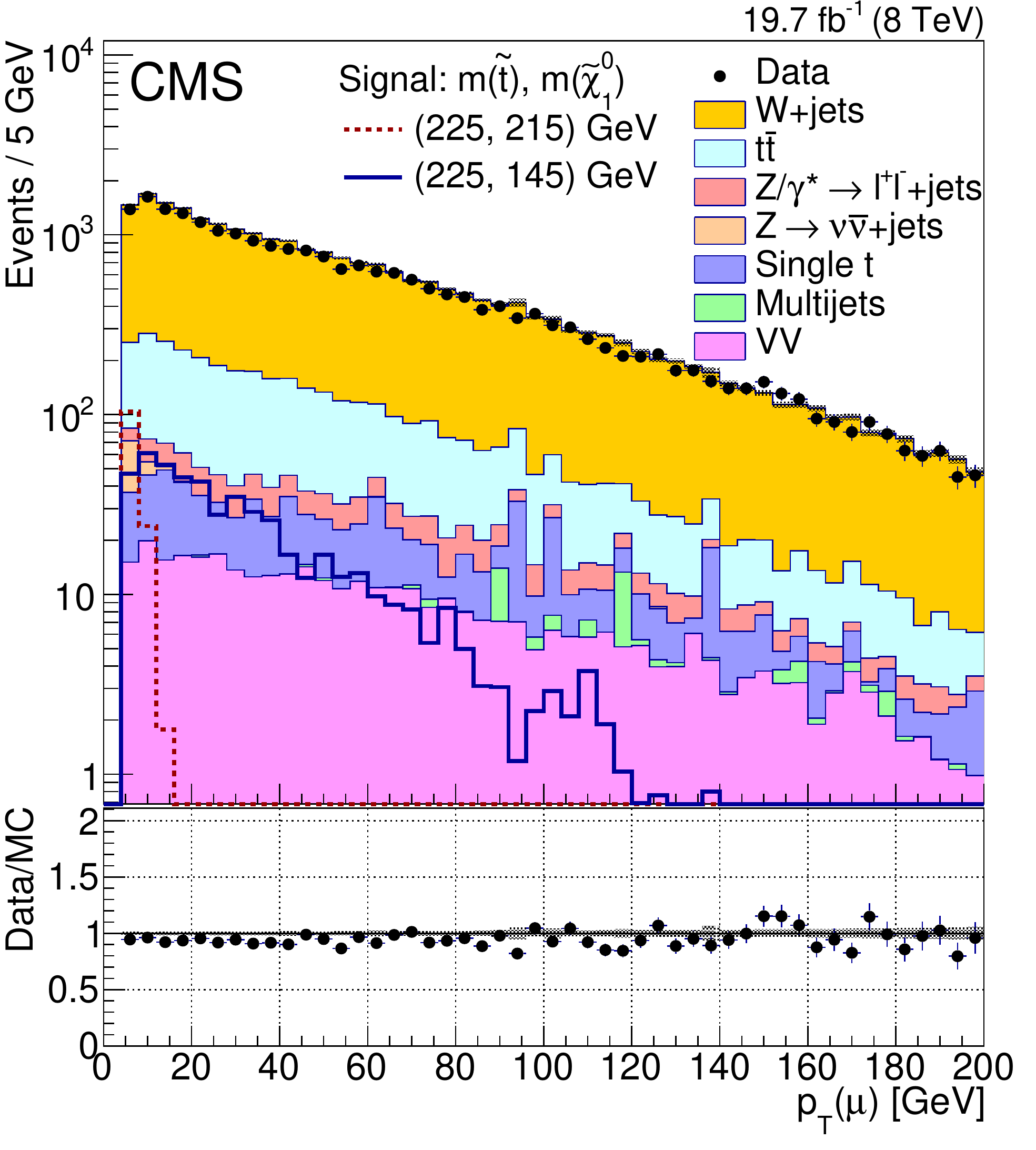} \hfil
  \includegraphics[width=.49\textwidth]{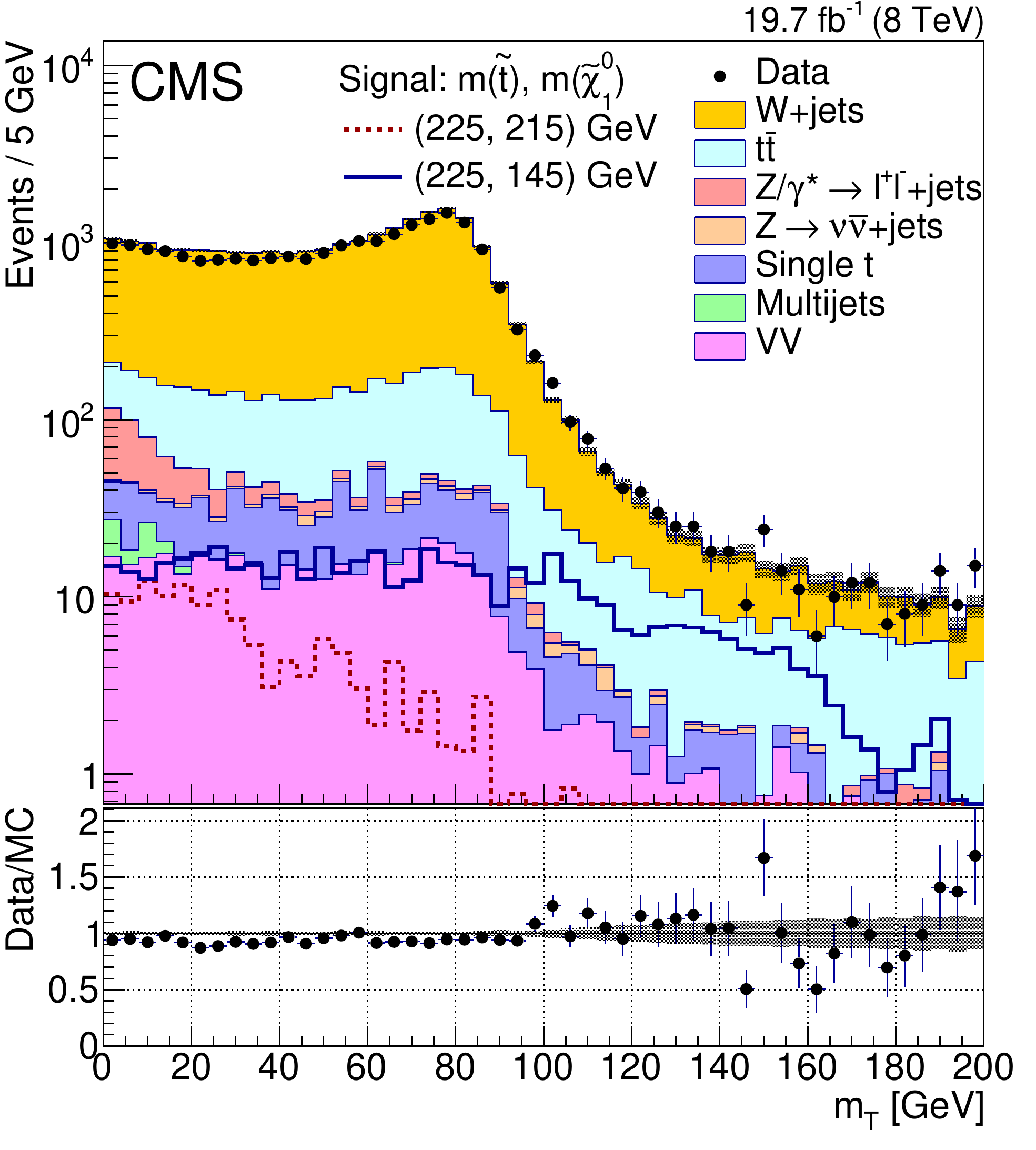}
\caption{Distributions of (\cmsLeft) muon \pt and (\cmsRight) \mt after the preselection of the single-muon analysis.
  For each plot, the variable shown has been excluded from the selection.
  Data are indicated by circles.
  The uncorrected background predictions from simulation are represented as filled, stacked histograms, and the shapes for two signal points with $\mass{\sTop}=225\GeV$ and mass splittings of $\DM=10$ and 80\GeV as dashed red and solid blue lines, respectively.
  The error bars and the dark, shaded bands indicate the statistical uncertainties of data and simulation, respectively.
  The lower panels show the ratio of data to the sum of the SM backgrounds.}\label{fig:slPresel}
\end{figure}

Distributions of the muon \pt and of \mt at this stage of the selection are presented in Fig.~\ref{fig:slPresel}.
They show good agreement between data and simulation.
The variation of the signal shapes is illustrated with two extreme cases of the mass splitting (10 and 80\GeV).

To maintain sensitivity over a large range of \DM values, several SRs are defined as listed in Table~\ref{tab:slSRCRdefs}.
Since signal leptons have low \pt, we impose an upper limit of $\pt<30\GeV$ in all these selections.
Because the muon \pt spectrum of the signal changes rapidly with \DM, the full range of muon \pt is subdivided into three bins in the calculation of the final results: 5--12, 12--20, and 20--30\GeV.

The signal region labelled as SRSL1 is designed for low values of \DM, where the \cPqb\ jets produced in the \sTop decays rarely pass the selection thresholds.
A veto on \cPqb-tagged jets strongly reduces the contribution from \ttbar events.
In addition, only events with negatively charged muons ($Q = -1$) are accepted, using the fact that the remaining \Wjets background shows significantly more positively than negatively charged muons~\cite{Chatrchyan:2014mua} while the signal is symmetric in the muon charge.
The acceptance for muons is reduced to the central region, $\abs{\eta}<1.5$, and the requirement on \CT is tightened to 300\GeV.
For signal points at low \DM, \mt is typically small, mainly due to the soft lepton \pt spectrum.
With increasing \DM, the average \mt increases and eventually the distribution extends to values above $\mass{\PW}$.
To cover the full range of \DM values, SRSL1 is therefore divided into three subregions, SRSL1a--c, defined by $\mt<60\GeV$, $60<\mt<88\GeV$, and $\mt>88\GeV$, respectively.

The second signal region (SRSL2) targets signals with higher mass splitting, where some of the \cPqb\ jets enter the acceptance.
Therefore, the \cPqb\ jet veto in the region $30<\pt<60\GeV$ is reversed, and at least one such jet is required.
Events with one or more \cPqb-tagged jets with $\pt>60\GeV$ are still rejected to reduce the \ttbar background.
In addition, the \pt threshold of the ISR jet candidate is raised to 325\GeV.
This second SR receives a strong contribution from \ttbar events.

\begin{table*}[tbh]
\centering
\topcaption{Definition of signal and control regions for the single-muon search.
For jets, the attributes ``soft'' and ``hard'' refer to the \pt ranges 30--60\GeV and $>60\GeV$, respectively.
For the calculation of the final results, each signal region (SRSL1a--c, SRSL2) is subdivided into three bins according to $\pt(\mu)$: 5--12, 12--20, and 20--30\GeV.}\label{tab:slSRCRdefs}
\centering
\begin{tabular}{lccc}
\hline
Variable & SRSL1a--c, CRSL1a--c & SRSL2, CRSL2 & CRSL(\ttbar) \\ \hline
\ETmiss (\GeV) & $>$300 & $>$300 & $>$200 \\
\HT (\GeV) & $>$400 & \NA & $>$300 \\
$\pt$(ISR jet) (\GeV) & $>$110 & $>$325 & $>$110 \\
Number of hard jets & $\leq$2 & $\leq$2 & $\leq$2 \\
$\Delta\phi$(hard jets) (rad) & $<$2.5 & $<$2.5 & $<$2.5 \\
\multirow{2}{*}{Number of \cPqb\ jets} & \multirow2{*}{0} & $\geq$1 soft & ($\geq$1 soft and $\geq$1 hard) \\
 &  & 0 hard & or ($\geq$2 hard) \\
$\pt(\mu)$ (\GeV) & 5--30 (SR), $>$30 (CR) & 5--30 (SR), $>$30 (CR) & $>$5 \\
$|\eta(\mu)|$ & $<$1.5 & $<$2.4 & $<$2.4 \\
$\dxy (\mu)$ (cm) & $<$0.02 & $<$0.02 & $<$0.02 \\
$\dz (\mu)$ (cm)  & $<$0.5 & $<$0.5 & $<$0.5 \\
$Q(\mu)$ & $-1$ & any & any \\
Lepton rejection & \multicolumn{3}{c}{no \Pe, $\tau$, or additional $\mu$ with $\pt>20\GeV$} \\
\mt (\GeVns{}) & $<$60 (a), 60--88 (b), $>$88 (c) &\NA & \NA \\
\hline
\end{tabular}
\end{table*}

\subsection{Background estimation}\label{sec:singleleptonBkg}

The following four background contributions are estimated by using data:
\Wjets and \ttbar production, which are the dominant components for the single-muon search;
$(\Z \to \nu\nu)+\text{jets}$, which is relevant for a signal region at high \mt as explained below;
and multijet production.
For the first three of these backgrounds, data/simulation scale factors are determined in suitable CRs and applied to the simulated yields in the SR.
The contribution of multijet events is estimated by using data only.
Rare backgrounds (other \Zg processes, and diboson and single top quark production) are predicted by using simulation.

Simulation provides only an imperfect description of the \pt spectrum for the main background samples (\Wjets, \ttbar).
Since the extrapolations from control to signal regions involve the lepton \pt spectrum, the \pt distributions of \PW\ bosons (for \Wjets events) and top quarks (for \ttbar events) are corrected based on measurements in data samples dominated by \ttbar, \Zjets, and \Wjets events before deriving the scale factors.

For the estimation of the \ttbar background, a single control region (CRSL(\ttbar)) is used:
events are required to pass the basic selection defined above and must include at least two \cPqb-tagged jets, with one of them in the \pt\ region above 60\GeV.
This CR has an estimated purity of 80\% in \ttbar events.
The observed event count in CRSL(\ttbar) is corrected for other background contributions and compared to the yield estimated from \ttbar simulation.
The resulting scale factor of 1.05 is then used to modify the predictions of the \ttbar simulation in all SRs.

The \Wjets yields from simulation are normalized in control regions associated to each of the four signal (sub-)regions SRSL1a--c (CRSL1a--c) and SRSL2 (CRSL2).
Control and signal regions differ only by the muon \pt range: in the CRs a muon with $\pt>30\GeV$ is required.
The control regions CRSL1a--c have an estimated purity of 80\% in \Wjets events.
For region CRSL2 this number is about 50\%, the remainder being dominated by \ttbar events.
Again, scale factors are derived after subtracting non-\Wjets backgrounds from the observed yields in the CR.
The \ttbar yields used in the subtraction are corrected by the scale factor determined as described in the previous paragraph.
The scale factors for \Wjets simulation vary from 0.88--1.18 in the four signal regions SRSL1a--c and SRSL2.

Each factor is applied to all three muon \pt bins of a signal region.
Systematic uncertainties are assigned related to the statistical uncertainties of the factors (6--30\%), and to the shape of the \pt spectrum as described later in this section.
The definitions of the single-lepton signal and control regions are summarized in Table~\ref{tab:slSRCRdefs}, and the expected compositions of the events in the control regions are shown in Table~\ref{tab:slCRs}.
For the benchmark signal models, the control regions would typically receive a contribution from signal events at the level of a few percent.
This effect is taken into account in the statistical analysis of the results.

\begin{table*}[tbh]
\centering
\topcaption{Contributions to the control regions of the single-muon analysis as determined from simulation before application of scale factors, together with the observed event counts.
All uncertainties are statistical.}\label{tab:slCRs}
\begin{tabular}{lyyyzz}
\hline
Background & \multicolumn{1}{c}{CRSL(\ttbar)} & \multicolumn{1}{c}{CRSL1a} & \multicolumn{1}{c}{CRSL1b} & \multicolumn{1}{c}{CRSL1c} & \multicolumn{1}{c}{CRSL2} \\ \hline
\Wjets & 67,9\pm3.6 & 323,3\pm6.4 & 141,9\pm4.3 & 30,3\pm2.0 & 36,5\pm2.3 \\
\ttbar & 471,0\pm9.6 & 19,5\pm2.2 & 9,9\pm1.5 & 6,1\pm1.2 & 37,5\pm3.5 \\
\Zgjets & 2,1\pm0.5 & 16,1\pm1.0 & 0,8\pm0.2 & 0,3\pm0.1 & 0,7\pm0.2 \\
\VV & 3,8\pm0.6 & 13,7\pm1.3 & 8,0\pm1.1 & 2,5\pm0.5 & 1,1\pm0.4 \\
Single top quark & 58,6\pm12.6 & 4,6\pm1.4 & 3,3\pm1.2 & 1,1\pm0.7 & 3,5\pm1.2 \\ \hline
Total SM & 603,4\pm16.2 & 377,1\pm7.1 & 165,0\pm4.8 & 40,3\pm2.5 & 79,4\pm4.3 \\ \hline
Data & \multicolumn{1}{l}{628} & \multicolumn{1}{l}{347} & \multicolumn{1}{l}{172} & \multicolumn{1}{l}{46} & \multicolumn{1}{l}{75} \\
\hline
\end{tabular}
\end{table*}

After applying the signal selection, with the exception of the requirement on muon \pt, the muon \pt spectra of \ttbar and \Wjets events are similar.
Therefore, the correction procedure leads to an anti-correlation of the estimates for the two categories and a relative uncertainty in the sum of the two contributions that is smaller than the uncertainty in a single component.
For this reason, the analysis is robust against variations in the relative yields of \ttbar and \Wjets events: a validation based on the direct estimation of the sum of both background components from the control regions CRSL1a--c and CRSL2 yields almost identical results in terms of the total background.
The anti-correlation between the two backgrounds is taken into account in the computation of the results described in Section~\ref{sec:results}.

The extrapolation of the correction factors from control to signal regions has been validated by comparing corrected yields from simulation to data in sideband regions.
Each of these sidebands is defined by one of the following changes with respect to the signal selection: (a) a lowering of the \ETmiss/\HT requirement to $200 < \CT <300\GeV$, (b) a change in the muon charge requirement, and (c) the condition of exactly one \cPqb-tagged jet with $\pt>60\GeV$.
The predictions in the sidebands are compatible with the observations, and the results are used to assign systematic uncertainties on the extrapolation of the scale factors to the SRs.
These uncertainties are 20\% for the estimate of the \ttbar background and 10--30\% for the estimate of the \Wjets background where the highest uncertainty applies to region SRL1c.

At high values of \mt, only a few \Wjets events pass the SRSL1c selection.
In this signal region, \Zjets production, with the \Z boson decaying to neutrinos, plus a nonprompt muon related to one of the jets, constitutes a non-negligible contribution.
This contribution is estimated from simulation, together with a correction derived from a data sample of events with two or more muons, selected by the single-muon trigger.
In this control sample, \Zjets events with \Z boson decays to muon pairs are used.
By using tighter muon selection criteria and restricting the mass of the dimuon system to be within 15\GeV of \mass{\Z}, a high-purity sample is obtained.
The events are used to mimic $\Z \to \nu\nu$ decays by removing the two daughter muons and adding their momenta to the \ETmiss vector.
The correction is applied as the product of two factors: $R_{\mu\mu}$, the inclusive data-to-simulation ratio, and $R_{\mu\mu\mu/\mu\mu}$, the ratio of the probabilities to observe a third, soft muon.
The first factor corrects the cross section in the $\mu\mu$ channel for a signal-like region.
Its measured value is $0.80 \pm 0.03$.
The double ratio $R_{\mu\mu\mu/\mu\mu}$ is determined in a looser selection to be $1.26 \pm 0.27$, yielding a total correction factor of $1.01 \pm 0.22$.
The uncertainties quoted above are statistical.
Systematic uncertainties due to the evolution with \ETmiss and \HT, or due to differences in the muon efficiency or acceptance between data and simulation, are negligible with respect to the statistical uncertainty.

The contribution from multijet events is estimated by inverting the requirements on muon isolation, the muon impact parameter, and the veto on leading jets in back-to-back configuration.
Assuming small correlations among the three variables mentioned above, the yield of multijet events can be estimated from the yield obtained with the fully inverted selection combined with the product of three reduction factors (one for each variable).
The estimated contributions to SRSL1 and SRSL2 are below 0.1 events and are therefore neglected.

A summary of the expected contributions of different background processes to the SRs is shown in Table~\ref{tab:slSRs} together with the yields of two benchmark signal points.

\begin{table*}[tbh]
\centering
\topcaption{Estimated background contributions for the signal regions of the single-muon analysis.
The scale factors determined in the control regions are applied.
For the signal samples, \mass{\sTop} and \mass{\PSGczDo} are shown in parentheses.
All uncertainties are statistical.}\label{tab:slSRs}
\begin{tabular}{lyzzz}
\hline
Background & \multicolumn{1}{c}{SRSL1a} &  \multicolumn{1}{c}{SRSL1b} &  \multicolumn{1}{c}{SRSL1c} &  \multicolumn{1}{c}{SRSL2} \\ \hline
\Wjets & 116,8\pm8.8 & 73,2\pm7.6 & 8,8\pm2.1 & 16,0\pm4.9 \\
\ttbar & 7,4\pm1.3 & 4,1\pm1.0 & 1,2\pm0.5 & 13,8\pm1.8 \\
$\cPZ \to \nu\nu$+jets & 1,1\pm0.4 & 1,2\pm0.4 & 1,5\pm0.5 & 0,3\pm0.2 \\
$\Zg \to \ell\ell$+jets & 4,4\pm0.5 & 0,2\pm0.1 & 0,2\pm0.1 & 0,5\pm0.2 \\
\VV  & 4,6\pm0.7 & 1,8\pm0.5 & 0,7\pm0.3 & 0,5\pm0.2 \\
Single top quark & 0,1\pm0.1 & 0,6\pm0.4 & \multicolumn{1}{c}{$<$0.3} & 1,0\pm0.7 \\ \hline
Total SM  & 134,5\pm8.9 & 81,3\pm7.8 & 12,3\pm2.3 & 32,1\pm5.3 \\ \hline
\rule{0pt}{1,05\vstop}\sTopsTop\ signal (250,230) & 32,5\pm2.8 & 6,2\pm1.2 & 4,7\pm1.0 & 7,1\pm1.3 \\
\rule{0pt}{1,05\vstop}\sTopsTop\ signal (300,250) & 11,0\pm1.0 & 4,2\pm0.6 & 5,1\pm0.7 & 10,7\pm1.0 \\
\hline
\end{tabular}
\end{table*}

\subsection{Background systematic uncertainties}\label{sec:singleleptonSyst}

In addition to the systematic uncertainties estimated in the previous subsections, the following systematic effects and associated uncertainties have been evaluated.

The full difference in the background estimates induced by the correction of the \pt spectrum of simulated \ttbar\ and \Wjets\ events is assigned as a systematic uncertainty.
The impact of the reweighting applied to \ttbar events is only significant for the signal region SRSL2, where the contribution of this background is the highest.

Changes in the polarization of the \PW\ boson can have an impact on the results since they change the balance between muon \pt and \ETmiss.
To quantify this effect, the polarization fractions $f_{\lambda=+1}$, $f_{\lambda=-1}$, and $f_{\lambda=0}$, associated with helicity $+1$, $-1$, and 0 amplitudes have been modified following three different scenarios: a 10\% variation of $f_{-1}-f_{+1}$ for both $\PW^{+}$ and $\PW^{-}$, a 5\% variation of $f_{-1}$, $f_{+1}$, and a 10\% variation of the longitudinal polarization fraction $f_{0}$ \cite{Chatrchyan:2011ig, ATLAS:2012au, Bern:2011ie}.

The uncertainties based on the comparison of data and simulation in the validation regions described in the previous subsection are propagated to the final estimate.
An uncertainty of 50\% is assigned to the cross sections of all non-leading backgrounds, including $\cPZ \to \nu\nu$, and propagated through the full estimation procedure.
An overview of all systematic uncertainties related to the background prediction is presented in Table~\ref{tab:slBkgSys}.
The dominant uncertainties are related to the limited statistical precision of the validation procedure and to the uncertainties in the shape of the muon \pt spectrum in \Wjets events.

\begin{table*}[tbh]
\centering
\topcaption{Relative systematic uncertainties in the background predictions in the signal regions of the single-muon search.
The labels refer to sources of systematic uncertainties discussed in Sections~\ref{sec:samples} and \ref{sec:singleleptonSyst}.
}\label{tab:slBkgSys}
\begin{tabular}{l....}
\hline
\multirow{2}{*}{Systematic effect} & \multicolumn{4}{c}{Uncertainty (\%)} \\ \cline{2-5}\\[-1.8ex]
                  &\multicolumn{1}{c}{SRSL1a} & \multicolumn{1}{c}{SRSL1b} & \multicolumn{1}{c}{SRSL1c} & \multicolumn{1}{c}{SRSL2} \\ \hline
Pileup & 0.5 & 0.8 & 3.0 & 0.5 \\
\PW\ \pt reweighting & 7.1 & 8.8 & 8.1 & 3.7 \\
\ttbar \pt reweighting & 0.8 & 0.5 & 0.1 & 5.4 \\
Jet energy scale & 2.4 & 3.2 & 2.1 & 6.0  \\
Jet energy resolution & 1.1 & 4.4 & 7.3 & 3.4 \\
\cPqb\ tagging & 0.1 & 0.1 & 0.5 & 1.3 \\
\PW\ polarization & 2.9 & 2.8 & 3.9 & 0.8 \\
Muon efficiency & 3.5 & 3.5 & 3.5 & 3.5 \\ \hline
\Wjets validation & 8.8 & 18.1 & 21.7 & 10.2  \\
\ttbar validation & 1.0 & 0.9 & 1.7 & 8.4 \\
Other backgrounds & 3.8 & 2.4 & 9.8 & 3.6 \\ \hline
Total uncertainty & 13.1 & 21.5 & 27.0 & 17.1 \\ \hline
\end{tabular}
\end{table*}

\section{Search in the dilepton channel}\label{sec:dilepton}

The analysis in the dilepton channel also starts from the common baseline selection described in Section~\ref{sec:samples}.
In this topology, less background is expected, and thus the selection
requirements on \ETmiss and the \pt of the ISR jet candidate are set
to be above 200 and 150\GeV, respectively, just above the trigger
thresholds.
To increase sensitivity, we select events in two signal regions defined by the \pt of the leading lepton: 5--15 and 15--25\GeV.
The second lepton is required to have $\pt<15\GeV$.
We require exactly two identified leptons of opposite sign, with at least one of them a muon.
Finally, events with an invariant mass of the dilepton pair $m(\ell\ell) < 5\GeV$ are rejected to remove a region that is difficult to simulate and to avoid any potential $\JPsi$ background.
Because the relative fraction of reconstructed leptons not arising from the decay of a \PW\ or \cPZ\ boson (``nonprompt'' leptons) is higher compared to the single-lepton channel, the isolation and identification criteria on the leptons are stricter.
On top of the muon identification used for the single-lepton topology, stricter requirements on the number of tracker hits, the quality of the track fit, and the match to signals in the muon detector are applied.
This selection is similar to the soft muon identification used for \cPqb-quark physics in CMS~\cite{Hermine}.
For electrons, the definitions for the $\PH\rightarrow\cPZ\cPZ\rightarrow4\ell$~\cite{Chatrchyan:2013mxa} analysis are used together with a stronger rejection of photon conversions.
For both flavours, the leptons are required to be isolated ($\Iabs < 5\GeV$ and $\Irel <0.5$) and to have impact parameter values \dxy and \dz smaller than 0.01\unit{cm}.
As in the region SRSL1 of the single-muon analysis, \cPqb-tagged jets are vetoed to suppress \ttbar backgrounds.
To remove potential multijet backgrounds, a selection on $\ETmiss/\HT>2/3$ is applied.

After this selection, one of the main backgrounds is \Zg production of $\tau$ pairs, with both $\tau$ leptons decaying leptonically.
Under the assumption that the direction of the reconstructed lepton is parallel to the $\tau$ direction, which is true to good approximation, the invariant mass of the $\tau$ pair, \mtautau, can be reconstructed by setting its transverse momentum equal to the hadronic recoil (the missing transverse momentum without the leptons).
All events with $\mtautau < 160\GeV$ are rejected.

The definitions of the dilepton signal (SRDL) and control (CRDL) regions are summarized in Table~\ref{tab:dlSRCRdefs}.

\begin{table*}[tbh]
\centering
\topcaption{Definition of signal and control regions for the dilepton search.
For the CRs, only changes with respect to the SR are shown.
Dashes indicate that no selection is applied.
For the lower limits on lepton \pt in the SR, the value used for electrons is shown in parentheses.
The SR is subdivided into two bins according to the \pt of the leading lepton: 5--15 and 15--25\GeV.
}\label{tab:dlSRCRdefs}
\begin{tabular}{l|c|cccccc} \hline
Variable & SRDL & \multicolumn{6}{c}{CRDL}\\
 & & \ttll &NPR1&NPR2&\VV&\cPZ&$\tau\tau$\\\hline
$Q(\ell_1) Q(\ell_2)$& $-1$ &  & $+1$ & $+1$ & & & \\
$\ell_1$ $\ell_2$ & $\mu\mu$,$\mu$\Pe,\Pe$\mu$ & $\mu\mu$,$\mu$\Pe & & $\mu\mu$,$\mu$\Pe & $\mu\mu$,$\mu$\Pe &  $\mu\mu$ & \\
$\pt (\ell_{1})$ (\GeVns{})     & 5(7)--25 & $>$25 & & $>$25   & $>$25 &  $>$125&\\
$\pt (\ell_{2})$ (\GeVns{})     & 5(7)--15 & $>$15 & & & $>$15 & $>$10 & \\
$\abs{\eta}(\ell)$   & $<$1.5 & & & &  & $<$2.1 &\\
\dxy, \dz $(\ell)$ (cm) & $<$0.01 &  & &  &  &$<$0.02, $<$0.5 & \\
\pt(ISR jet)  (\GeVns{}) & $>$150 & & & & & & \\
\pt(jet3) (\GeVns{})  & $<$60 & & & & & & \\
Number of \cPqb\ jets & 0 & 1 & & & 0 (loose id.) & &  \\
Number of jets            & $\ge$1 & & & & 1 or 2 & & \\
$|\Delta\phi(\ell_1,\text{ISR jet})|$ (rad) & \NA & & & & $>$1 & & \\
\ETmiss (\GeVns{}) & $>$200 & $>$125 & & $>$125 & $>$125 & \NA & \\
\ETmiss/\HT & $>$2/3 & \NA & & \NA & \NA & \NA & \\
\LT (\GeVns{})         & \NA   & $>$225 & & $>$225 & $>$225 & \NA & \\
$\LT/\HT$   & \NA   & $>$2/3 & & $>$2/3 & $>$2/3 & & \\
$\pt(\mu\mu)$ (\GeVns{}) & \NA  &   & & & & $>$200 & \\
$\pt(\mu\mu)/\HT$& \NA &  & & & & $>$2/3 & \\
$m(\ell\ell)$ (\GeVns{}) & $>$5 & & & & $>$50 & $>$10& \\
$m(\tau\tau)$ (\GeVns{}) & $>$160 & & & & & \NA &$<$160 \\ \hline
\end{tabular}
\end{table*}

\subsection{Background prediction} \label{sec:Backgrounds_Predictions}
Four different background categories are predicted from data: dileptonic \ttbar events (\ttll, $\ell:e\mu\tau$), which constitute the largest background; diboson production such as \PW\PW\ or \PW\cPZ\ (the second-largest background); and \Zg production of $\tau$ pairs with leptonic $\tau$ decays.
Backgrounds with one nonprompt lepton, \ie\ \Wjets and semileptonic \ttbar events (\ttl), are the fourth category.
Half of the background events contain at least one $\tau$ lepton that decays leptonically.
The negligible ($\approx$1\%) contribution of rare processes ($\ttbar\cmsSymbolFace{V}$, $\ttbar\PH$, \cPqt\PW, and $\Wpm\Wpm$) is predicted by using simulation.
For each of the four categories, a CR enriched in such processes is defined in data, from which we derive correction factors to correct yields from simulation.

In all CRs, the requirements on jets are the same as in the SR.
Several CRs use events with higher lepton \pt compared to the SR.
In these regions, the leading lepton has to be a muon, and events are selected by using the single-muon trigger described before.
The relative lepton isolation has to be smaller than 0.12 and the muon identification criteria are tightened.
Apart from the \Zg control region, the \ETmiss requirement is lowered to 125\GeV and the \ETmiss selection of the signal region is instead applied to \LT, the sum of \ETmiss and the \pt of the leading lepton. The present selection is $\LT>225\GeV$ to take into account that for the default selection \LT is also up to 25\GeV higher than \ETmiss.
In this way, the event yields in the CRs can be increased  while maintaining kinematics similar to the SR even in the presence of a higher-\pt lepton.

To achieve a clean control sample of dileptonic \ttbar events (CRDL(\ttll)), we require exactly one \cPqb-tagged jet.
This jet must not be the leading jet to ensure a distribution in \pt of the \ttbar system similar to that in the SR.
We require one muon with $\pt>25\GeV$ and a subleading lepton with $\pt>15\GeV$.
Backgrounds other than \ttll are subtracted from data before calculating the ratio between data and the prediction from simulation for \ttll in the CR.
This ratio is used to rescale the simulated \ttll yields in the SR.

For the CR enriched in nonprompt leptons (CRDL(NPR)), we use the union of two samples.
The first sample (CRDL(NPR1)) corresponds to the SR with the exception that the leptons are required to have the same charge.
It was checked that in the selected kinematic region, the origins for NPR leptons, mainly heavy quarks, occur at a similar fraction as in the SR.
In addition, the kinematics of these nonprompt leptons is very similar in signal and control regions.
For the second sample (CRDL(NPR2)), same-sign events with a leading lepton \pt above 25\GeV are used, and the CR selection of $\ETmiss > 125\GeV$, $\LT>125\GeV$, and $\LT/\HT>2/3$ is applied.
Under these conditions the origins and kinematics of the nonprompt leptons are similar between signal and control regions, since the NPR contribution in the signal region is mostly related to the subleading lepton.
Again the data yield in the combined CR is corrected for other backgrounds, such as diboson events, by using simulation.
The ratio of the corrected yield to the simulated NPR yield in the CR is used to rescale the simulated NPR yield in the SR.

For the prediction of \Zg events, two separate CRs are defined.
The first one is used to correct for any effects on \mtautau (CRDL(\cPZ)).
For this purpose, a clean sample of \Zg events with decays to a pair of muons is used.
The invariant mass of the muon pair has to be higher than 10\GeV and the \ETmiss selection is applied to the \pt of the muon pair.
Three bins are defined as a function of this momentum: 200--300, 300--400, and $>$400\GeV.
We use the reconstructed muon pair \pt
to measure the resolution of the hadronic recoil along and perpendicular to the direction given by the muon pair both in data and simulation.
The resulting scaling factors of the recoil resolution are applied to the simulation to recompute the efficiency of the \mtautau selection in the SR.
A second control region (CRDL($\tau\tau$)) is used to measure in data the probability of $\Zg \to \tau\tau$ events leading to two soft leptons and very high \ETmiss.
To do so, we use the SRDL selection with the requirement on \mtautau inverted to $<160\GeV$.
After subtracting other backgrounds in this region by using simulation, the observed yield is multiplied by the corrected \mtautau efficiency to predict the number of \Zg events in the SR.

For the diboson control region (CRDL(\VV)) one muon with $\pt>25\GeV$ is required.
The \pt of the second lepton has to be ${>} 15\GeV$.
To further enhance the diboson fraction and reduce the otherwise dominant \ttbar background, at most two jets are allowed, events with a jet passing a looser working point of the \cPqb\ tagging algorithm are rejected, and the azimuthal angle between the leading lepton and the leading jet has to be $>1\unit{rad}$.
Finally, we require $m_{\ell\ell}$ to be above 50\GeV.
Contributions of \ttll, NPR, and \Zg to CRDL(\VV) are estimated with methods similar to those used for the SR.
Backgrounds due to rare processes are subtracted by using the simulation.
After this correction, the ratio of the number of data to simulated diboson events is built and used to rescale the simulated \VV yield in the SR.

The background contribution from multijet events is negligible in our final selection. Apart from the fact that we require
high \ETmiss and two leptons, we also select $\ETmiss/\HT > 2/3$ to reject any residual multijet events.
To evaluate the efficacy of this selection, a test was performed by inverting this requirement to have a region that
should have significant multijet background if there were any. For this region, data yields were compared
with the simulation results  of all considered background categories (which do not include multijet events) and were found to be in agreement. Further tests were performed by using
the electron-electron channel and by relaxing the upper limits on \dxy and \dz to 0.05\unit{cm},
which showed no indication for a contamination by multijet events. These tests confirmed that, as expected, we can assume that multijet background
is negligible in our final selection that employs much tighter requirements against multijet events.

The event yields for data and simulation in the different CRs that are the basis for the scale factors applied in the SRs are shown in Table~\ref{tab:YIELDS_h_CR}.
The predicted event yields per background for each search bin are presented in Table ~\ref{tab:YIELDS_h_SR_OS_0b}.
The impact of a potential signal contamination is found to be only relevant for control regions CRDL(NPR) and CRDL($\tau\tau$), with an effect of a few percent on the total background prediction in the signal regions, and is taken into account in the statistical analysis of the results.

\begin{table*}[tbh]
\centering
\topcaption{Contributions to the control regions of the dilepton analysis as expected from
simulation before application of scale factors, together with the observed event counts.
All uncertainties are statistical.}\label{tab:YIELDS_h_CR}
\begin{tabular}{lyzzZ} \hline
Background & \multicolumn{1}{c}{\rule{0pt}{1.05\vstop}CRDL(\ttll)}  & \multicolumn{1}{c}{CRDL(NPR)}  &  \multicolumn{1}{c}{CRDL(\VV)} & \multicolumn{1}{c}{CRDL($\tau\tau$)} \\ \hline
\rule{0pt}{1.05\vstop}\ttll &      119,1\pm2.4 &  0,27\pm0.11 & 30,3\pm1.2 &   0,15\pm0.08 \\
\rule{0pt}{1.05\vstop}\ttl  &      1,09\pm0.29 &  4,7\pm0.6   &  0,30\pm0.14&  0,11\pm0.11 \\
\Wjets           &      \multicolumn{1}{c}{$<$0.4} & 3,4\pm1.3 & \multicolumn{1}{c}{$<$0.4}  &  0,7\pm0.7 \\
\Zgjets          &      0,4 \pm0.4 &  \multicolumn{1}{c}{$<$0.30}        &  4,9\pm1.3 &   2,8\pm0.9 \\
\VV              &      2,4 \pm0.6 &   0,62\pm0.11 & 45,9\pm1.8 &   0,13\pm0.09 \\
Rare backgrounds &      14,9\pm2.7 &   1,0\pm0.5   &  6,4\pm1.7 &   \multicolumn{1}{c}{$<$0.21} \\ \hline
Total SM background &   138,0\pm3.7&  10,0\pm1.5   & 87,8\pm3.0 &   3,9\pm1.1 \\ \hline
Data             &      \multicolumn{1}{l}{119}          &   \multicolumn{1}{l}{11}             &  \multicolumn{1}{l}{8}           &    \multicolumn{1}{l}{5} \\ \hline
\end{tabular}\end{table*}

\begin{table*}[bht]
\centering
\topcaption{Estimated background contributions for the two signal regions of the dilepton search.
The scale factors determined in the control regions are applied.
For the signal samples, \mass{\sTop} and \mass{\PSGczDo} are shown in parentheses.
All uncertainties are statistical.}\label{tab:YIELDS_h_SR_OS_0b}
\begin{tabular}{lZZZ} \hline
Background &  \multicolumn{1}{c}{$\pt(\ell_{1})$: 5--15\GeV}  &  \multicolumn{1}{c}{$\pt(\ell_{1})$: 15--25\GeV} & \multicolumn{1}{c}{Inclusive}\\ \hline
\rule{0pt}{1.05\vstop}\ttll &     0,75 \pm0.19 &  2,08 \pm0.37 &  2,8 \pm0.4 \\
\rule{0pt}{1.05\vstop}\ttl ,\Wjets  & 0,60 \pm0.33 &  1,3 \pm0.7 &  1,9 \pm0.8 \\
\Zgjets           & \multicolumn{1}{c}{$<$0.30} &  0,5 \pm0.5 &  0,5 \pm0.5 \\
\VV               & 0,74 \pm0.27 & 1,6 \pm0.5 &  2,4 \pm0.5 \\
Rare backgrounds  & 0,03 \pm0.01 & 0,08 \pm0.04 &  0,11 \pm0.04 \\ \hline
Total SM background & 2,1 \pm0.5 & 5,6  \pm1.0  &  7,7  \pm1.1 \\ \hline
\rule{0pt}{1.05\vstop}\sTopsTop\ signal (225,145)  & 4,2  \pm1.2  & 6,3  \pm1.6  & 10,4 \pm2.0\\
\rule{0pt}{1,05\vstop}\sTopsTop\ signal (300,250)  & 4,0 \pm0.6 & 3,8 \pm0.6 &  7,8\pm0.8 \\ \hline
\end{tabular}
\end{table*}

To test the prediction methods, we define several validation regions that are enriched in specific backgrounds but expected to be free of signal.
The first region is equivalent to the signal region, except for an inversion of the veto on \cPqb-tagged jets. This region is used to test the prediction of low-\pt leptons in \ttll events.
The next validation region is identical to CRDL(\ttll), except that the \pt of the subleading lepton is required to be below 15\GeV. This region provides a further test of the prediction of the soft-lepton rate.
Another validation region is the same as CRDL(\VV), apart from the fact that all selections used to enrich the region in diboson events are inverted.
In addition, a validation region that has a composition in backgrounds similar to the signal region is defined. For this, one muon with \pt above 25\GeV is required, while the second lepton must be soft ($\pt<15\GeV$). All validation regions show reasonable agreement between prediction and observation.

\subsection{Background systematic uncertainties}

In addition to the common uncertainties from object reconstruction and simulation as described in Sections~\ref{sec:objects} and \ref{sec:samples}, the following systematic uncertainties that are specific to the individual background predictions are considered.

In estimating the \ttll background, the polarization of the \PW\ boson resulting from the top quark decay is varied.
In addition, the spin correlation between the two top quarks is changed by 20\%, since this might affect how often both leptons are soft~\cite{Aad:2014mfk}.
As in the single-lepton channel, the difference due to the reweighting of the top quark \pt spectrum in \ttbar simulation is taken as a further uncertainty.
However, its effect is small due to the background prediction method used.

For a conservative assessment of the uncertainty related to the estimate of NPR backgrounds, the fractions of leptons from \cPqb\ and \cPqc\ hadrons are varied by 50\% and 100\%, respectively.
The relative fraction of  $\ttbar$ to \Wjets events is altered by rescaling both contributions by $\pm50\%$.
The largest of these variations is used as the uncertainty.
Furthermore, the \pt and $\abs{\eta}$ distributions in the CR are varied to reflect potential residual differences in the kinematics between signal and control regions.
Moreover, the polarization of the \PW\ boson is varied to estimate the uncertainty due to polarization modelling.

The cross sections for \PW\PW~\cite{ATLAS:2012mec, Khachatryan:2015sga} and also \PW\cPZ\ and \cPZ\cPZ~\cite{Chatrchyan:2014aqa} production have been measured at the LHC, and both the total and differential cross sections show reasonable agreement between data and simulation.
To estimate the uncertainties related to \VV production, the polarization of the vector bosons is altered by 10\%, as well as the fraction of the diboson pair momentum that a single boson carries.
In addition, the cross section corresponding to events with low $\mass{\gamma^*}$ between 5 and 12\GeV is varied by 100\% to account for any potential shape mismodelling of the dilepton mass.

In the estimation of the \Zg background, the effect of the recoil resolution correction is used to derive an uncertainty due to a potential mismodelling of the resolution.
The cross section of rare processes is varied by $\pm$50\% throughout the analysis (also in the CRs), and the effect is propagated to the event yields in the SR.

\begin{table*}[hbt]
 \centering
 \topcaption{Relative systematic uncertainties in the background predictions in the signal regions of the dilepton search.}\label{tab:2Lsystematics}
 \begin{tabular}{l..} \hline
\multirow{2}{*}{Systematic effect}              &  \multicolumn{2}{c}{Uncertainty (\%)} \\ \cline{2-3}\\[-1.8ex]
                               &  \multicolumn{1}{c}{$\pt(\ell_{1})$: 5--15\GeV}  &  \multicolumn{1}{c}{$\pt(\ell_{1})$: 15--25\GeV}  \\ \hline
Statistical uncertainty & 21.9 &  18.3 \\ \hline
Jet energy scale        & 1.0  &   2.8 \\
\cPqb\ tagging           & 1.5  &   1.4 \\
Electron efficiency     & 1.3  &   1.1 \\
Muon efficiency         & 6.0  &   4.5 \\ \hline
\rule{0pt}{1.05\vstop}$\ttbar$ background    & 5.1  &   5.4 \\
NPR background          &10.1  &   5.6 \\
\Zg  background         & \multicolumn{1}{c}{$<$0.1}  &   2.3 \\
\VV background         & 8.0  &   2.6 \\
Rare backgrounds & 3.7 & 3.3 \\ \hline
Total uncertainty & 26.9 & 21.1 \\ \hline
\end{tabular}
\end{table*}

A summary of all uncertainties can be found in Table~\ref{tab:2Lsystematics}.
The dominating uncertainty stems from the limited number of simulated events with nonprompt leptons in the SRs.
\section{Results and interpretation}\label{sec:results}

The observations and background predictions for the signal regions of the single-lepton and dilepton searches are summarized in Table~\ref{tab:results}.
Observed and predicted yields are in good agreement and give no indication of the presence of signal.

The modified-frequentist CL$_\mathrm{S}$ method~\cite{frequentist_limit,Read:2002hq,LHC-HCG} with a one-sided profile likelihood ratio test statistic is used to define 95\% confidence level (CL) upper limits on the production cross section as a function of the sparticle masses.
Statistical uncertainties related to the observed number of events in CRs are modelled as Poisson distributions.
All other uncertainties are assumed to be multiplicative and are modelled with log-normal distributions.
The impact of a potential signal contamination in the control regions is taken into account in the calculation of the limits for each signal point.

\begin{table*}
\centering
\topcaption{Summary of observed and expected background yields in the signal regions of the single-lepton and dilepton searches.
The uncertainties in the background yields include statistical and systematic contributions.
Transverse momenta are shown in units of \GeV.}\label{tab:results}
\begin{tabular}{cccccc|ccc}
\hline
 \multicolumn{6}{c|}{Single muon} & \multicolumn{3}{c}{Dilepton} \\
 \pt($\mu$) & & SRSL1a &  SRSL1b &  SRSL1c &  SRSL2 & \pt($\ell_1$) & & SRDL \\ \hline
 5--12 & exp.  &       41.1 $\pm$ 6.3
   &       29.7 $\pm$ 7.2
   &       4.3 $\pm$ 1.5
   &       11.3 $\pm$ 2.9
   & 5--15 & exp. &  2.1 $\pm$ 0.6 \\
 & obs. &       42
   &       17
   &       3
   &       16
   &  & obs. & 2 \\
 12--20 & exp. &       44.2 $\pm$ 6.8
  &       25.1 $\pm$ 6.2
  &       3.1 $\pm$ 1.2
  &       8.5 $\pm$ 2.4
  & 15--25 & exp.  &      5.6 $\pm$ 1.2 \\
 & obs. &       39
  &       14
  &       4
  &       16
  &  & obs. & 4 \\
 20--30 & exp.  &       49.2 $\pm$ 7.5
  &       26.5 $\pm$ 6.5
  &       5.0 $\pm$ 1.8
  &       12.2 $\pm$ 3.0 & \multicolumn{3}{c}{} \\
 & obs. &       40
  &       28
  &       5
  &       9 & \multicolumn{3}{c}{} \\ \hline
 All & exp. &   134.5 $\pm$ 19.8
  &       81.3 $\pm$ 19.1
  &       12.3 $\pm$ 4.0
  &       32.1 $\pm$ 7.7
  & All & exp. & 7.7 $\pm$ 1.4  \\
 & obs. &       121
  &       59
  &       12
  &       41
  & & obs. & 6 \\ \hline
\end{tabular}
\end{table*}

Systematic uncertainties in the signal yields related to the determination of the integrated luminosity~\cite{CMS-PAS-LUM-13-001} (2.6\%), pileup ($\approx$2\%), energy scales~\cite{CMS-PAS-JME-13-004,Khachatryan:2014gga} (up to 7\%), object identification efficiencies~\cite{Khachatryan:2015hwa,Chatrchyan:2012xi} (up to 10\%), and uncertainties in the parton distribution functions~\cite{Alekhin:2011sk,Botje:2011sn,Ball:2012cx,Martin:2009iq,Lai:2010vv} (up to 6\%) and the modelling of ISR~\cite{Chatrchyan:2013xna} ($\approx$20\%) have been evaluated.
Correlations between the systematic uncertainties in different signal regions are taken into account, where applicable.
All systematic uncertainties are treated as nuisance parameters in the calculation of the limits, with the exception of the theoretical uncertainty on the inclusive SUSY production cross section.
The latter is shown in the form of an up- and downward variation of the observed mass limits.

\begin{figure}
\centering
\includegraphics[width=0.49\textwidth]{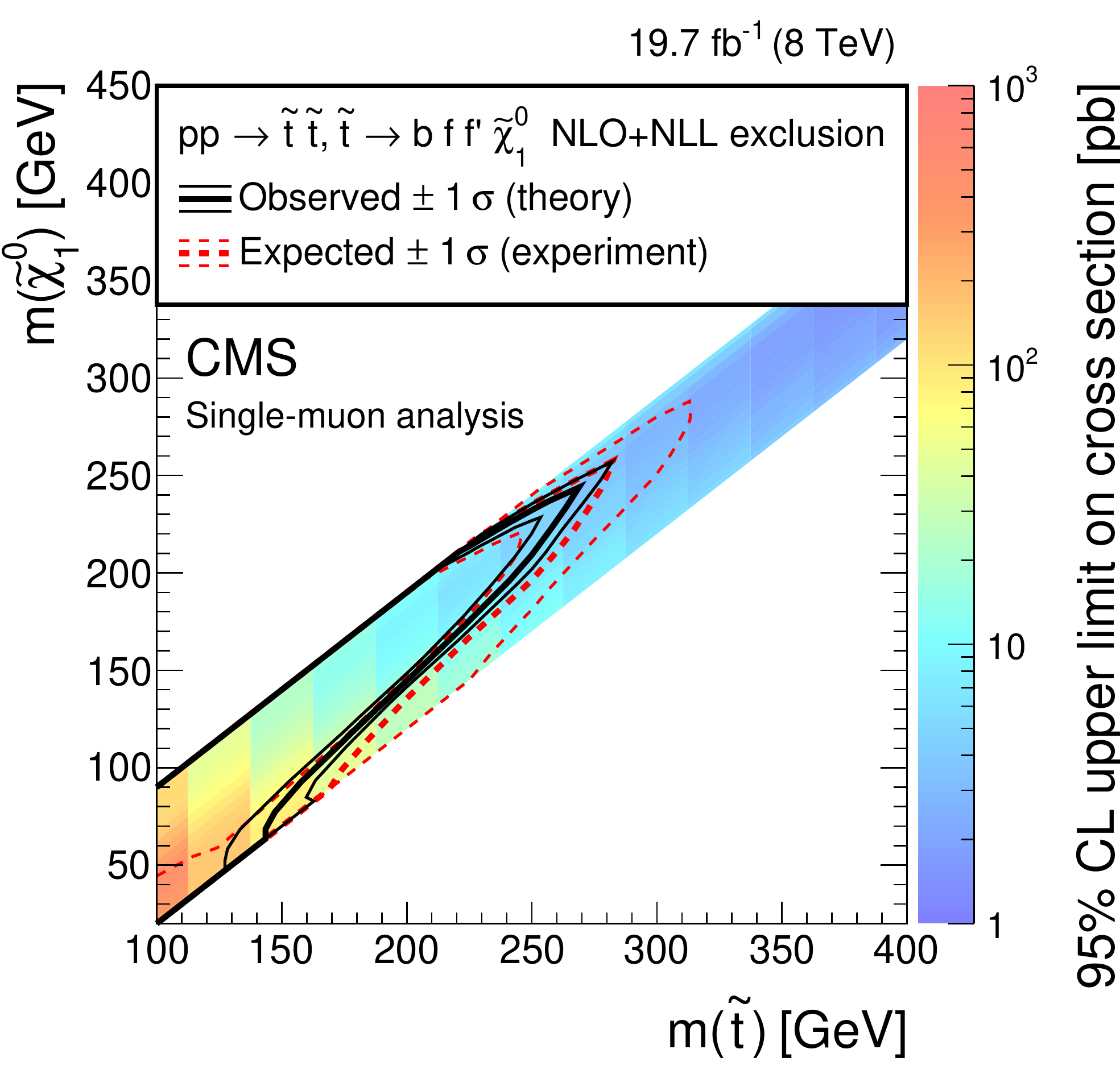} \hfil
\includegraphics[width=0.49\textwidth]{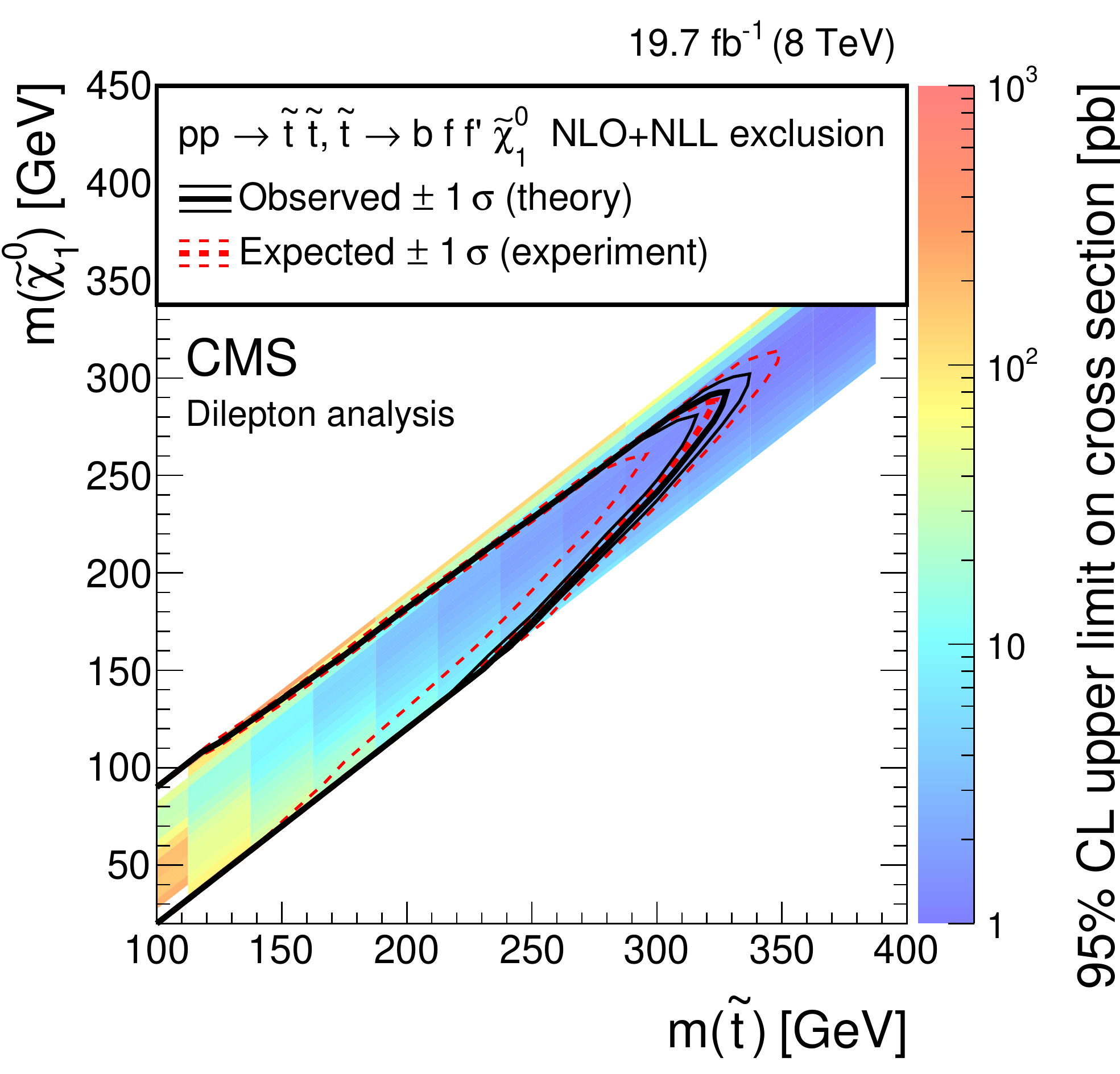}
\caption{
Cross section and mass limits at 95\% CL in the \mass{\PSGczDo} and \mass{\sTop} mass plane for the (\cmsLeft) single-muon and (\cmsRight) dilepton searches.
The colour shading corresponds to the observed limit on the cross section.
The solid (dashed) lines show the observed (expected) mass limits, with the thick lines representing the central value and the thin lines the variations due to the theoretical (experimental) uncertainties.
}\label{fig:stopLimits}
\end{figure}

The limits obtained for top squark pair production in the single-muon and the dilepton searches are shown in Fig.~\ref{fig:stopLimits} \cmsLeft and \cmsRight, respectively, under the assumption of a 100\% branching fraction of the four-body decay.
By using the \sTop\ pair production cross section calculated at next-to-leading order (NLO) + next-to-leading logarithm (NLL) precision~\cite{bib-nlo-nll-01,bib-nlo-nll-02,bib-nlo-nll-03,bib-nlo-nll-04,bib-nlo-nll-05}, the cross section limits can be converted into excluded regions in the \sTop--\PSGczDo\ mass plane.
Uncertainties in these cross sections are determined as detailed in Ref.~\cite{Kramer:2012bx}.
At $\DM = 25\GeV$, the dilepton search excludes \sTop masses below 316\GeV.
Here and in the following all quoted values for mass limits conservatively refer to the ${-}1\sigma$ variation of the predicted cross section.
The single-muon search shows a smaller reach in \mass{\sTop} ($\approx$250\GeV) but has a higher sensitivity at the lowest considered mass splitting of 10\GeV, where values up to $\approx$210\GeV are excluded.
In the intermediate \DM region ($\approx$20--70\GeV), these results considerably extend existing limits~\cite{Aad:2014kra,Aad:2014nra,Aad:2015pfx}.
They are complementary to the results of searches in the monojet topology~\cite{Khachatryan:2015wza,Aad:2014kra,Aad:2015pfx}.

\begin{figure}
\centering
\includegraphics[width=\cmsFigWidth]{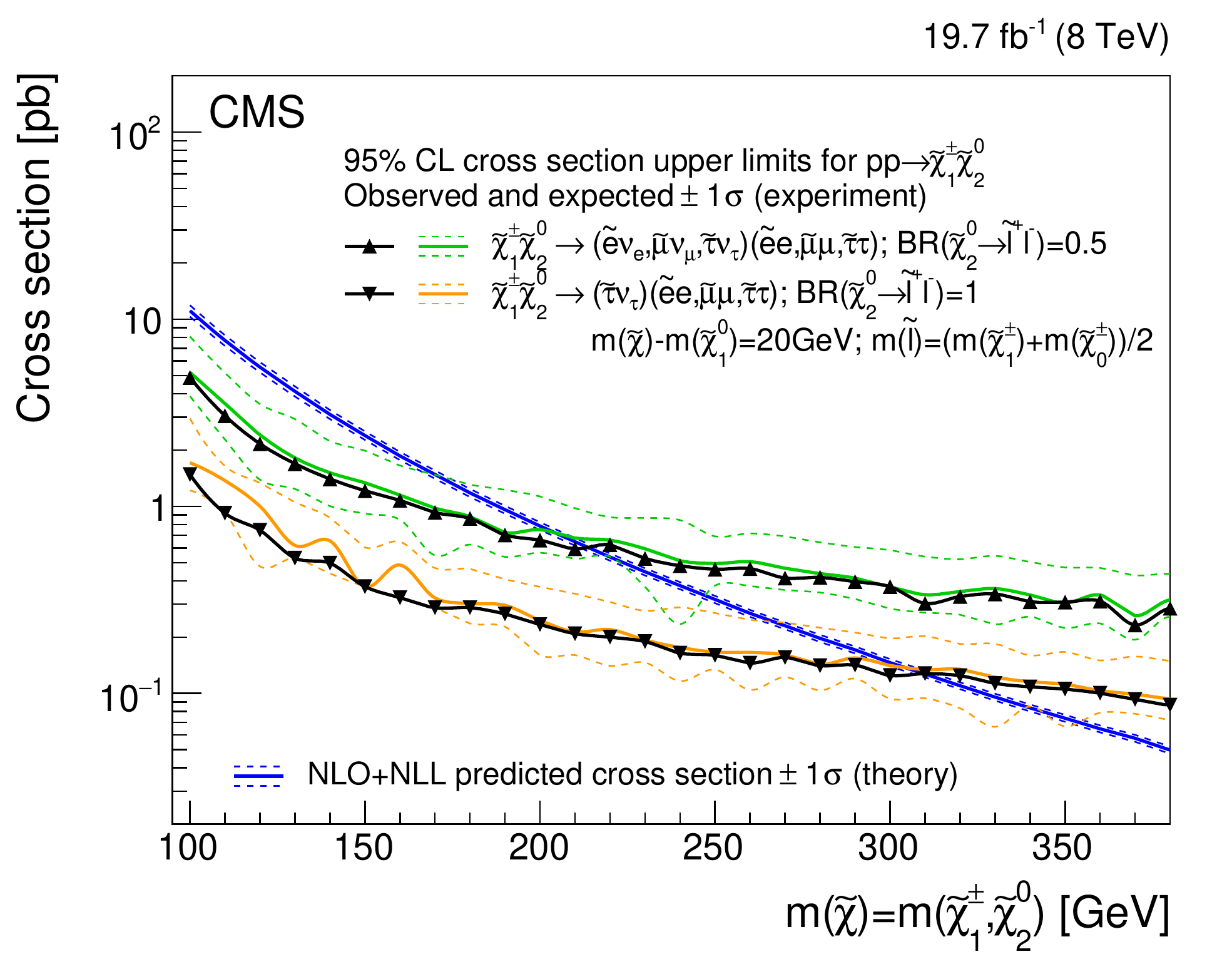}
\caption{
Cross section limits at 95\% CL obtained from the search in the dilepton channel as a function of the common \PSGcpmDo/\PSGczDo mass.
The black lines with symbols correspond to the observed limit, while the solid and dashed coloured lines represent the expected limit and the ${\pm}1\sigma$ bands corresponding to the experimental uncertainties, respectively.
The flavour-democratic ($\tau$-enriched) cases of the model are indicated by green (orange) lines and upward- (downward-) pointing triangular symbols.
The solid and dashed blue lines without symbols correspond to the predicted cross section for chargino-neutralino production and its uncertainties.
}\label{fig:chiLimits}
\end{figure}

In the case of chargino-neutralino pair production, the results of the dilepton analysis are used.
For the model involving decay chains with sleptons, the slepton masses are set to $(\mass{\PSGczDo}+\mass{\PSGcpDo})/2$.
Instead of using branching fractions derived from complete SUSY models, two extreme decay scenarios are studied in order to illustrate the dependence on the final state.
In the ``flavour-democratic" scenario, both the neutralino and the chargino would decay via the supersymmetric partners of the left-handed leptons (\sLl) and of the neutrinos (\sNu) with equal branching fractions to all lepton flavours.
In this scenario, the fraction of events with at least two charged leptons is reduced by 50\% due to the $\PSGczDt \to \nu\nu\PSGczDo$  decay channel.
In the ``$\tau$-enriched" scenario, the decays would proceed via the supersymmetric partners of the right-handed leptons.
In this case, the decay $\PSGczDt \to \nu\nu\PSGczDo$ is not present, and we assume equal branching fractions of the \PSGczDt\ into the three charged lepton flavours and exclusive decays of the chargino to $\tau$ leptons.
In Fig.~\ref{fig:chiLimits}, the 95\% CL cross section limits are presented for a mass splitting of $\DM \equiv \mass{\PSGc} - \mass{\PSGczDo} = 20\GeV$.
Comparing with the predicted cross section, calculated at NLO+NLL
precision with the
\textsc{Resummino}~\cite{Fuks:2012qx,Fuks:2013vua,Fuks:2013lya} program,
95\% CL limits on $\mass{\PSGc}$ of 212 and 307\GeV are obtained for the flavour-democratic and $\tau$-enriched scenarios, respectively.
In these compressed scenarios, the new limits slightly improve current results~\cite{Khachatryan:2014qwa,Aad:2015eda} in the flavour-democratic scenario and exceed them by $\approx$200\GeV for the $\tau$-enriched scenario.
As for the latter case, the dominant decays lead to final states with opposite-sign leptons.

\section{Summary}\label{sec:conclusions}

A search for supersymmetry with compressed mass spectra is performed in events with soft leptons, moderate to high values of  \ETmiss, and one or two hard jets, compatible with the emission of initial-state radiation.
The data sample corresponds to 19.7\fbinv of proton-proton collisions recorded by the CMS experiment at $\sqrt{s} = 8\TeV$.
Two event categories are considered: events with a single, soft muon and events in which a second, soft electron or muon is present.

The first target of this search is the pair production of top squarks with a mass splitting of at most 80\GeV with respect to the LSP.
At low mass splitting, lepton momenta are low, and the \cPqb~jets do not enter the acceptance.
At higher values of \DM, the average lepton momentum increases and soft \cPqb~jets can be reconstructed.
Therefore, signal regions are further divided according to the \pt of the leading lepton and the presence or absence of a soft \cPqb-tagged jet.
In the single-lepton search the transverse mass of the lepton-\ETmiss system is used as an additional discriminant.

The main backgrounds to this search are \Wjets and \ttbar production.
Contributions to the signal regions from these and several nonleading background sources are estimated by using data in control regions to normalize the simulated yields.
These estimates are tested with data in validation regions.

The observations in the signal regions are compatible with the SM background predictions.
In the absence of any indication of signal, cross section limits are set at 95\% CL in the \sTop--\PSGczDo\ mass plane.
These results are used to extract mass limits based on a reference cross section for top squark pair production and assuming a 100\% branching fraction for the four-body decay $\sTop \to \cPqb \fermion \fermion' \PSGczDo$.
The most stringent limit on the mass of the top squark is obtained in the dilepton channel: $m(\sTop) > 316\GeV$ at 95\% CL for a mass splitting of 25\GeV.
These results extend existing limits in the four-body decay channel of the top squark~\cite{Aad:2014kra,Aad:2014nra,Aad:2015pfx} and complement the analyses performed in the $\sTop \to \cPqc \PSGczDo$ channel~\cite{Khachatryan:2015wza,Aad:2015pfx}.

The results obtained in the dilepton channel are also used to set limits on models of chargino-neutralino production in a compressed spectrum with a mass difference between \PSGczDt/\PSGcpDo and \PSGczDo of 20\GeV.
Based on the 95\% CL upper limit on the cross section in the case of flavour-democratic leptonic decays of these particles, a lower limit on the common \PSGcpDo/\PSGczDt mass is set at 212\GeV.
If chargino decays proceed exclusively via the $\tau$ channel, and in the absence of the $\PSGczDt \to \sNu\nu$ decay mode, this limit increases to 307\GeV, well above existing limits~\cite{Khachatryan:2014qwa,Aad:2015eda}.

\begin{acknowledgments}
We congratulate our colleagues in the CERN accelerator departments for the excellent performance of the LHC and thank the technical and administrative staffs at CERN and at other CMS institutes for their contributions to the success of the CMS effort. In addition, we gratefully acknowledge the computing centres and personnel of the Worldwide LHC Computing Grid for delivering so effectively the computing infrastructure essential to our analyses. Finally, we acknowledge the enduring support for the construction and operation of the LHC and the CMS detector provided by the following funding agencies: BMWFW and FWF (Austria); FNRS and FWO (Belgium); CNPq, CAPES, FAPERJ, and FAPESP (Brazil); MES (Bulgaria); CERN; CAS, MoST, and NSFC (China); COLCIENCIAS (Colombia); MSES and CSF (Croatia); RPF (Cyprus); MoER, ERC IUT and ERDF (Estonia); Academy of Finland, MEC, and HIP (Finland); CEA and CNRS/IN2P3 (France); BMBF, DFG, and HGF (Germany); GSRT (Greece); OTKA and NIH (Hungary); DAE and DST (India); IPM (Iran); SFI (Ireland); INFN (Italy); MSIP and NRF (Republic of Korea); LAS (Lithuania); MOE and UM (Malaysia); CINVESTAV, CONACYT, SEP, and UASLP-FAI (Mexico); MBIE (New Zealand); PAEC (Pakistan); MSHE and NSC (Poland); FCT (Portugal); JINR (Dubna); MON, RosAtom, RAS, and RFBR (Russia); MESTD (Serbia); SEIDI and CPAN (Spain); Swiss Funding Agencies (Switzerland); MST (Taipei); ThEPCenter, IPST, STAR and NSTDA (Thailand); TUBITAK and TAEK (Turkey); NASU and SFFR (Ukraine); STFC (United Kingdom); and DOE and NSF (USA).

Individuals have received support from the Marie-Curie programme and the European Research Council and EPLANET (European Union); the Leventis Foundation; the A. P. Sloan Foundation; the Alexander von Humboldt Foundation; the Belgian Federal Science Policy Office; the Fonds pour la Formation \`a la Recherche dans l'Industrie et dans l'Agriculture (FRIA-Belgium); the Agentschap voor Innovatie door Wetenschap en Technologie (IWT-Belgium); the Ministry of Education, Youth and Sports (MEYS) of the Czech Republic; the Council of Science and Industrial Research, India; the HOMING PLUS programme of the Foundation for Polish Science, cofinanced from European Union, Regional Development Fund; the OPUS programme of the National Science Center (Poland); the Compagnia di San Paolo (Torino); MIUR project 20108T4XTM (Italy); the Thalis and Aristeia programmes cofinanced by EU-ESF and the Greek NSRF; the National Priorities Research Program by Qatar National Research Fund; the Rachadapisek Sompot Fund for Postdoctoral Fellowship, Chulalongkorn University (Thailand); the Chulalongkorn Academic into Its 2nd Century Project Advancement Project (Thailand); and the Welch Foundation, contract C-1845.
\end{acknowledgments}
\bibliography{auto_generated}

\cleardoublepage \appendix\section{The CMS Collaboration \label{app:collab}}\begin{sloppypar}\hyphenpenalty=5000\widowpenalty=500\clubpenalty=5000\textbf{Yerevan Physics Institute,  Yerevan,  Armenia}\\*[0pt]
V.~Khachatryan, A.M.~Sirunyan, A.~Tumasyan
\vskip\cmsinstskip
\textbf{Institut f\"{u}r Hochenergiephysik der OeAW,  Wien,  Austria}\\*[0pt]
W.~Adam, E.~Asilar, T.~Bergauer, J.~Brandstetter, E.~Brondolin, M.~Dragicevic, J.~Er\"{o}, M.~Flechl, M.~Friedl, R.~Fr\"{u}hwirth\cmsAuthorMark{1}, V.M.~Ghete, C.~Hartl, N.~H\"{o}rmann, J.~Hrubec, M.~Jeitler\cmsAuthorMark{1}, V.~Kn\"{u}nz, A.~K\"{o}nig, M.~Krammer\cmsAuthorMark{1}, I.~Kr\"{a}tschmer, D.~Liko, T.~Matsushita, I.~Mikulec, D.~Rabady\cmsAuthorMark{2}, B.~Rahbaran, H.~Rohringer, J.~Schieck\cmsAuthorMark{1}, R.~Sch\"{o}fbeck, J.~Strauss, W.~Treberer-Treberspurg, W.~Waltenberger, C.-E.~Wulz\cmsAuthorMark{1}
% ...
\vskip\cmsinstskip
% ...
1:~~Also at Vienna University of Technology, Vienna, Austria\\
2:~~Also at CERN, European Organization for Nuclear Research, Geneva, Switzerland\\
% ...

\end{sloppypar}
\end{document}